\documentclass[12pt]{article}

\usepackage{geometry}
\makeatletter
\ifx\@undefined\href
  \def\href#1#2{#2}
\fi
\makeatother

\newcommand{\nit}{\noindent}
\newcommand{\nl}{\newline}

\newcommand{\bg}{\beta}
\newcommand{\gam}{\gamma}

\newcommand{\lb}{\lambda}

\newcommand{\sg}{\sigma}

\newcommand{\og}{\omega}

\newcommand{\lh}{\left(}
\newcommand{\rh}{\right)}

\newcommand{\vs}[1]{\vspace{#1 em}}

\newcommand{\pl}{\partial}

\begin{document} 
\begin{flushright}
NIKHEF 99-003\\
DAMTP-1999-17\\
hep-th/9901163\\
Januari 26, 1999
\end{flushright}
\vskip 5em
\begin{center}
{\Large\bf An index theorem for non-standard Dirac operators}\\[2em]
{\sc\large Jan-Willem van Holten,  Andrew Waldron}\\[.5em]
NIKHEF\\
P.O. Box 41882\\
1009 DB Amsterdam\\
The Netherlands\\[2em]
{\sc\large Kasper Peeters}\\[.5em] 
DAMTP, Cambridge University\\
Silver Street\\
Cambridge CB3 9EW\\
United Kingdom\\[2em]
{\tt t}$(8\cdot n)${\tt @nikhef.nl},~~ $n=4,3,2$
\end{center}
\vskip 2em
\begin{quote}
\begin{center}
\bf Abstract
\end{center}

On manifolds with non-trivial Killing tensors admitting a square 
root of the Killing-Yano type one can construct non-standard Dirac 
operators which differ from, but commute with, the standard Dirac 
operator. We relate the index problem for the non-standard Dirac
operator to that of the standard Dirac operator. This necessitates 
a study of manifolds with torsion and boundary and we summarize
recent results obtained for such manifolds.
\end{quote}

\newpage

On manifolds like the four-dimensional Kerr-Newman and Taub-NUT 
manifolds, the geodesic equations are integrable because of the 
existence of a symmetric second-rank Killing tensor 
$K^{\mu\nu}$~\cite{cart2}, allowing the construction of a constant of motion 
\begin{equation} 
\label{1}
K = \frac{1}{2}\, K^{\mu\nu} p_{\mu} p_{\nu}\, .
\end{equation}
The Killing tensor condition
\begin{equation}
\label{2}
K_{\lh \mu\nu;\lb\rh} = 0
\end{equation}
is actually equivalent with the conservation of $K$, {\it i.e.}~$K$ 
commutes with the worldline Hamiltonian 
\begin{equation}
H = \frac{1}{2}\, g^{\mu\nu} p_{\mu} p_{\nu} 
\label{3}
\end{equation} 
in the sense of Poisson brackets: 
\begin{equation}
\left\{ K, H \right\} = 0\, . 
\label{4}
\end{equation} 
Related to this, the Klein-Gordon equation with minimal 
electromagnetic coupling on these background spaces is soluble 
by separation of variables~\cite{cart3}. Making use of the observation 
of Penrose and Floyd that such Killing tensors can have a square 
root of Killing-Yano-type~\cite{penr1,floy1}:
\begin{equation}
K_{\mu\nu} = f_{\mu\lb} f^{\lb}_{\;\;\nu}\, , 
\label{5}
\end{equation}
with the properties
\begin{equation}
f_{\mu\nu} = - f_{\nu\mu}\, , \hspace{3em} 
f_{\mu\nu;\lb} + f_{\mu\lb;\nu} = 0\, , 
\label{6}
\end{equation}
Carter and McLenaghan showed the existence of a Dirac-type linear 
differential operator which commutes with the standard Dirac 
operator~\cite{cart4}. Such non-standard Dirac operators take the form 
\begin{equation} 
D_f \equiv \gam_5 \gam^{\lb} \lh f_{\lb}^{\;\mu} D_{\mu} - 
  \frac{1}{3!} \sg^{\mu\nu} H_{\mu\nu\lb} \rh\, , 
\label{7}
\end{equation} 
where (in contrast to the more familiar Dirac operators associated 
to covariantly constant complex structures) the second term is nonzero,
\begin{equation}
H_{\mu\nu\lb} = f_{\left[ \mu\nu; \lb \right]} = f_{\mu\nu; \lb}\, .
\label{8}
\end{equation}
The square brackets in the middle expression denote complete 
anti-symmetrization with unit weight; the last equality is obtained 
straightforwardly from the Killing-Yano conditions (\ref{6}). 
In this letter we discuss the zero mode spectrum of such Dirac
operators (see \cite{diet1,diet2} for a systematic classification of
manifolds admitting solutions to~(\ref{6})).
\bigskip

\noindent The covariant derivative $D_{\mu}$ includes the standard 
spin-connection term: 
\begin{equation}
D_{\mu} = \pl_{\mu} - \frac{1}{2} \og_{\mu}^{\;\;ab} \sg_{ab}\, , 
\label{9}
\end{equation}
with the $\sg_{ab}$ the usual Dirac representation of the 
generators of the Lorentz group on spinors. In the following 
we frequently use the representation of vectors and tensors in 
local Lorentz components, obtained by contraction with a vielbein 
$e_{\mu}^{\;a}$ or the inverse vielbein $e_a^{\;\mu}$, {\it e.g.}
\begin{equation}
f_{\mu}^{\;a} = f_{\mu}^{\;\nu} e_{\nu}^{\;a}\, , \hspace{2em} 
  c_{abc} = e_a^{\;\mu} e_b^{\;\nu} e_c^{\;\lb} H_{\mu\nu\lb}\, .
\label{10}
\end{equation} 
In this notation one can write the Dirac operator (\ref{7}) as 
\begin{equation}
D_f = \gam_5 \gam^{a} \lh f_a^{\;\mu} D_{\mu} - \frac{1}{3!} 
  \sg^{bc} c_{abc} \rh\, . 
\label{11}
\end{equation}
The standard Dirac operator (without torsion) is 
\begin{equation} 
D = -i \gam^{a} e_a^{\;\mu} D_{\mu}\, ; 
\label{12}
\end{equation}
then the Killing-Yano properties of $f_a^{\;\mu}$ guarantee the 
commutation relation 
\begin{equation} 
\left[ D_f, D \right] = 0\, . 
\label{13}
\end{equation} 
Thus, the standard and non-standard Dirac operators can be 
diagonalized simultaneously, which is at the root of 
Chandrasekhar's observation that the Dirac equation in the 
Kerr-Newman solution is separable~\cite{chan7}. We stress that although one
often formulates the commutation relation~(\ref{13}) in terms of
Poisson brackets (for example, in classical studies of spinning
particles), when examining index problems it is essential that one
considers quantum commutators.
\vs{1} 

\nit
In even-dimensional spaces one can define the index of a Dirac
operator as the difference in the number of linearly independent 
zero modes with eigenvalue +1 and $-1$ under $\gam_5$: 
\begin{equation}
{\rm index} \lh D \rh = n^0_+ - n^0_-\, . 
\label{14}
\end{equation} 
The index is useful as a tool to investigate topological properties 
of the base-space, as well as in computing anomalies in quantum 
field theory; for a review see {\it e.g.}~\cite{alva12,alva17}. Eq.(\ref{13}) 
now
leads to a simple but remarkable result for the index of the 
non-standard Dirac operator $D_f$, to wit the \nl 
{\em theorem:}
\begin{equation}
{\rm index} \lh D_f \rh = {\rm index} \lh D \rh\, . 
\label{15}
\end{equation} 
Below we sketch the proof of this theorem and discuss some of its 
possible implications. 

Let $\left| \lb, \mu \rangle \right.$ denote an orthonormal basis 
of simultaneous eigenvectors of the standard and non-standard 
Dirac operators: 
\begin{equation} 
D \left| \lb, \mu \rangle \right. = \lb \left| \lb, \mu \rangle 
 \right.,  \hspace{3em}
D_f \left| \lb, \mu \rangle \right. = \mu \left| \lb, \mu \rangle 
 \right..
\label{16}
\end{equation} 
Now as $\gam_5$ anti-commutes with each of the Dirac operators, 
all non-zero eigenvalues occur in pairs of opposite signs: 
\begin{equation} 
D \gam_5\, \left| \lb, \mu \rangle \right.\, =\, -\, \gam_5 D\, 
 \left| \lb, \mu \rangle \right. \, =\, -\, \lb \gam_5\, 
 \left| \lb, \mu \rangle \right., 
\label{17}
\end{equation} 
and similarly for $D_f$. Projecting the states of fixed 
eigenvalue onto the eigenstates of $\gam_5$:
\begin{equation} 
\left| \lb,\mu; \pm \rangle \right.\, \equiv\, \frac{1}{2}\, 
  \lh 1 \pm \gam_5 \rh\, \left| \lb,\mu \rangle \right., 
\label{18}
\end{equation} 
one obtains the equivalent results
\begin{equation}
\begin{array}{l}
\gam_5\, \left| \lb,\mu; \pm \rangle \right.\, =\, \pm\, 
 \left| \lb,\mu; \pm \rangle \right., \\
\\
D\, \left| \lb,\mu; \pm \rangle \right. \, =\, \lb\, 
 \left| \lb,\mu; \mp \rangle \right., \hspace{2em} 
D_f\, \left| \lb,\mu; \pm \rangle \right. \, =\, \mu\, 
 \left| \lb,\mu; \mp \rangle \right.. 
\end{array}
\label{19}
\end{equation} 
On the other hand, for the zero-modes a mismatch between positive 
and negative chirality states may occur. Note however, that even 
in the kernel of the Dirac operator $D$ the pairing of chirality 
eigenstates holds for those vectors which are non-zero modes of 
$D_f$: 
\begin{equation} 
D_f\, \left| 0,\mu; \pm \rangle \right.\, =\, \mu\, \left| 0, \mu;
  \mp \rangle \right.. 
\label{20} 
\end{equation} 
Therefore the only states in the kernel of $D$ which can contribute 
to the index ${\rm index} (D)$ are the {\em simultaneous} zero modes of
$D$ and $D_f$, {\it i.e.}~the states $\left| 0, 0; \pm \rangle\right.$. 
Denoting the number of positive, resp.\ negative, chirality 
double-zero modes by $n_{\pm}^{(0,0)}$, the index of the standard 
Dirac operator is 
\begin{equation}
{\rm index} \lh D \rh\, =\, n^{(0,0)}_+\, -\, n^{(0,0)}_-\, . 
\label{21} 
\end{equation}
Now the symmetry of the algebra of (anti-)commutation relations of 
$D$, $D_f$ and $\gam_5$ under interchange of $D$ and $D_f$ implies, 
that again only the simultaneous zero-modes of $D_f$ and $D$ 
contribute to the index of the non-standard Dirac operator $D_f$; 
hence 
\begin{equation}
{\rm index} \lh D_f \rh\, =\, n^{(0,0)}_+\, -\, n^{(0,0)}_-\, 
                   =\, {\rm index} \lh D \rh\, . 
\label{22}
\end{equation}
This is the result we set out to prove. 
\vs{1}

\nit
We now comment on some interesting consequences of this theorem. 
It is well-known~\cite{alva12,alva17} that the index of a Dirac operator can 
be computed using path-integral methods, being identical to the 
Witten index of a supersymmetric quantum mechanical model for a 
spinning particle with supercharge 
\begin{equation}
Q\, =\, e_a^{\; \mu} \Pi_{\mu} \psi^{a}\, . 
\label{23}
\end{equation}
Here $\psi^{a}$ is the anti-commuting spin variable of the particle, 
forming a $d =1$ supermultiplet with the base-space co-ordinates 
$x^{\mu}$~\cite{riet3,riet5}, and $\Pi_{\mu}$ is the covariant momentum of 
the particle, which is related to the canonical momentum $p_{\mu}$ 
by 
\begin{equation}
\Pi_{\mu} = p_{\mu} + \frac{i}{2} \og_{\mu ab} \psi^{a} \psi^{b}\, . 
\label{24}
\end{equation} 
The existence of a Killing-Yano tensor and a non-standard Dirac
operator on the base-manifold now manifests itself as a new  
non-standard supersymmetry~\cite{gibb1} with supercharge 
\begin{equation} 
Q_f\, =\, f_a^{\; \mu} \Pi_{\mu} \psi^{a}\, +\, \frac{i}{3!}\, 
  c_{abc} \psi^{a} \psi^{b} \psi^{c}\, . 
\label{25} 
\end{equation} 
The main algebraic properties of the supercharges under 
Dirac-Poisson brackets are 
\begin{equation}
 \left\{ Q, Q \right\}\, =\, -2i H\, ,
 \hspace{2em} \left\{ Q_f, Q \right\}\, =\, 0\, , \hspace{2em}  
 \left\{ Q_f, Q_f \right\}\, =\, - 2i K\, ,
\label{26}
\end{equation} 
where $K$ now represents the supersymmetric extension of the 
Killing constant (\ref{2}); for example, in the case of Kerr-Newman 
space-time it is the supersymmetric 
extension of Carter's constant~\cite{gibb1}. Let us again stress that
for the study of the index, where one is interested in the spectrum of states,
one must consider the quantization of the spinning particle model. This 
problem is taken up in detail in~\cite{kas_index}

One may expect that the equality of the indices of the standard 
and non-standard Dirac operators now translates to the equality of 
the Witten index for the standard and non-standard supersymmetry. 
However, this translation involves some subtleties, which we 
briefly discuss here. First, let us recall that the Witten index 
is defined as the difference between the number of odd and even 
fermion number zero modes of the Hamiltonian of a supersymmetric 
theory, which may be computed in regularized form as~\cite{alva12,alva17}
\begin{equation}
{\rm index}_w = \lim_{\bg \rightarrow 0}\, \mbox{Tr} 
  \lh (-1)^F e^{-\bg H} \rh\, . 
\label{27}
\end{equation} 
Here $H$ is the Hamiltonian, obtained as the square of the 
standard supercharge. The formula in (27) makes only definite sense once 
one understands
what is meant by the symbol ``Tr''. Of course one should really view
this as a sum over the spectrum of the regulator however, even then
three distinct cases can occur~\cite{akho1}. (I) The spectrum
is discrete, in which case all states save for the zero modes cancel
pairwise and the index is some $\beta$-independent integer counting the
disparity between positive and negative 
chirality zero modes. (II) The spectrum includes a continuum separated
from the (discrete) zero mode sector by a gap. Then, so long as one can
show that the densities of positive and negative non-zero modes are
equal, the result is again an integer independent of $\beta$. (III)
The spectrum includes a continuum not separated from zero in which 
case the index is no longer guaranteed to be integer nor
$\beta$-independent. For the case we are interested in, namely the
chiral gravitational anomaly on manifolds with boundary, Atiyah, Patodi
and Singer~\cite{atiy1} have shown how to impose boundary conditions for
spinors which ensure a well-posed index problem (case (I) in fact).
Essentially their non-local boundary conditions stipulate the spectrum
of the bulk modes in terms of the discrete spectrum of the compact
boundary manifold (see~\cite{kas_index,atiy1,nino1,roem1,eguc3} for further 
details).

Similarly, a quantity for the non-standard supersymmetry can be obtained by  
defining the regularized expression 
\begin{equation}
{\rm index}_f = \lim_{\bg \rightarrow 0}\, \mbox{Tr} 
  \lh (-1)^F e^{-\bg K}\rh\, .
\label{28}
\end{equation} 
The equality of the indices then is equivalent to the statement 
that taking the limit $\bg \rightarrow 0$ these two expressions are 
equal: 
\begin{equation}
{\rm index}_f = {\rm index}_w\, . 
\label{29}
\end{equation} 
Comparison of the two expressions shows, that ${\rm index}_f$ actually 
is to be interpreted as the Witten index of a theory in which 
$K$ is the Hamiltonian. The relation between the theories of
which $H$, resp.\ $K$ are the Hamiltonians was investigated in 
detail in~\cite{riet2}. Here we recall in particular two results 
of that paper: 
\begin{enumerate}
\item If $K$ is a Killing-Carter constant of motion with respect to 
the Hamiltonian $H$, then $H$ is a constant of motion in the theory 
with Hamiltonian $K$. \\
{\em Corollary:} if $K_{\mu\nu}$ is a 
symmetric Killing tensor on a manifold with metric $g_{\mu\nu}$, 
then $g_{\mu\nu}$ is a symmetric Killing tensor on a manifold 
with metric $K_{\mu\nu}$. Following~\cite{riet2} we call this 
reciprocal relation between manifolds with metrics and Killing 
tensors interchanged a {\em Killing duality.} 
\item The correspondence can be extended to supercharges in the 
following way: if $Q_f$ is a non-standard Killing-Yano supercharge 
in a theory with standard supercharge $Q$ and Hamiltonian $H$, 
then $Q$ is a Killing-Yano supercharge in a dual theory with 
supercharge $Q_f$ and Hamiltonian $K$; but if $H$ and $K$ are 
different then at least one of the manifolds is endowed with 
torsion. This is because of the inclusion of the totally 
anti-symmetric tensor $c_{abc}$ in the supercharge $Q_f$, as 
well as in the corresponding non-standard Dirac operator $D_f$. 
\end{enumerate}

\nit
In order to implement the definition (\ref{28}) and prove the 
equality (\ref{29}) it is therefore necessary to study index 
theorems on manifolds with torsion. Furthermore, typically the
manifolds in question ({\it e.g.}~Taub--NUT and Kerr--Newman) 
have a boundary. Index theorems on manifolds with
boundary are the subject of the Atiyah--Patodi--Singer (APS) index
theorem. The extension of their work to include torsion has recently
been given in~\cite{kas_index}.
This is a rather  intricate subject. In particular, if one wishes to
regard the Dirac operators $D$ and $D_f$ as operators on independent 
manifolds then one has to carefully study the Hilbert spaces in which they
act. These and other detailed issues have been handled in depth
in~\cite{kas_index} so here we simply provide a summary of the most important
results.
\vs{1}

\nit
On manifolds with boundary the index splits into three terms. The first
is the bulk contribution which can be obtained by a heat kernel, 
Pauli--Villars or
supersymmetric path integral approach. This term has been independently
computed by
several groups~\cite{obuk2,doba1,kas_index} and the result is
\begin{equation}
\label{bulk} 
{\rm index}({\rm bulk})\,=\,
\frac{1}{24.8\pi^2}\int_{{\cal M}}\,\Big[
 R( e)^{mn}\wedge R( e)_{nm}
- \frac{1}{2} F(A)\wedge F(A) -2\, \sqrt{g}\, D_\mu {\cal K}^\mu
\Big] \, .
\end{equation}
Here $F(A)$ is the abelian field strength of the one form obtained by 
dualising the totally antisymmetric part of the torsion (remember that
the Dirac operator couples only to the trace and totally antisymmetric
parts of the contortion tensor). The vector ${\cal K}^\mu$ is given by
\begin{equation}
\label{kay}
{\cal K}^\mu = \Big( D^{\nu} D_{\nu} + \frac{1}{4}  A^\nu  A_\nu + 
\frac{1}{2} R\Big)\, 
 A^\mu \quad \, .
\end{equation} 
The second term required for the index is a boundary correction term
given by the APS $\eta$ invariant. The idea is quite simple, when the
manifold is of product form near the boundary ({\it i.e.}~a cylinder
with the boundary manifold as cross section) the solutions of the
Dirac equation can be derived directly from the solutions of the boundary
Dirac operator. Therefore APS have proven that the additional
correction can be constructed directly from the eigenvalues of the
boundary Dirac operator
\begin{equation}
\label{boundary}
\mbox{index(boundary)}=-\frac{1}{2}\Big(\eta(0)+h\Big)\, ,
\end{equation} 
where $h$ denotes the number of zero modes of the boundary Dirac 
operator and the $\eta$ invariant is given by
\begin{equation}
\eta(s)=\sum_{\{l\neq0\}}\frac{{\rm sign}(l)}{|l|^s}\,
\end{equation}
with the sum running over the set of all non-vanishing eigenvalues
$\{l\neq0\}$ of the boundary Dirac operator. Observe that $\eta(0)$
essentially counts the disparity between the number of positive and
negative boundary eigenvalues. Typically $\eta(0)$ must be computed on
a case by case basis, however for boundary manifolds given by a 
squashed
$S^3$ 
this computation has been performed some time ago by
Hitchin~\cite{hitc2}. The generalization of this computation to the
torsion-full case may again be found in~\cite{kas_index}.
\bigskip

\nit Of course, one is not always so lucky enough to have a manifold
approaching a cylinder near the boundary so that in the most general
case a third correction is necessary. In the torsion-free case this
correction was found by Gilkey~\cite{gilk1} and amounts to the
boundary integral over the difference between the Chern--Simons form
of the product manifold obtained by extending the boundary
three-manifold to a cylinder and the Chern--Simons form of the actual
four-manifold in question.  In the torsion-full case,
in~\cite{kas_index} it is shown how to extract a generalized
Chern--Simons from the integrand of~(\ref{bulk}).  Denoting quantities
computed on the product manifold by the sharp symbol, the generalized
Chern--Simons correction to the index is given by
\begin{equation}
\label{chern}
\mbox{index(CS correction)} = \frac{1}{24.8\pi^2}\int_{\partial{\cal M}} 
\Big[ C^\sharp(A)
-2\,{\cal K}^\sharp-
C(A)+2\,{\cal K} \Big]\, .
\end{equation}
Note that ${\cal K}$ denotes the three form obtain by dualising the
vector ${\cal K}^\mu$ in~(\ref{kay}).
\bigskip

\nit Finally, it is enlightening to see all these corrections computed
in an explicit example. The case of Taub--NUT and its torsion-full
dual have been analysed in~\cite{kas_index}. For the Taub--NUT
manifold with metric
\begin{equation}
\label{taub--nut}
{\rm d}s^2 =\frac{r+2m}{r}\,\Big[ {\rm d}r^2 + r^2\, {\rm d}\theta^2 +
    r^2\sin^2\theta\, {\rm d}\phi^2\Big]+ \frac{4r m^2}{r+2m}\,\Big[
    {\rm d}\psi + \cos\theta\, {\rm d}\phi\Big]^2\, ,
\end{equation}
the bulk contribution was calculated to be $1/12$ by
Hawking~\cite{hawk1}. There is no clash with the expectation of
integer counting of zero modes by the index since the boundary
corrections of APS (as computed by Hitchin) and Gilkey were shown to
exactly cancel the bulk term~\cite{roem1,eguc4,pope1} so that the
final result for the index of the Taub--NUT manifold is exactly zero.
Therefore, we would expect, from the considerations above, also a
vanishing result for the index of the dual manifold with metric given
by
\begin{equation}
\label{e:dual_metric}
{\rm d}s^2_f = 
    \frac{r+2m}{r}\Big[ {\rm d}r^2
   +\frac{m^2r^2}{(r+m)^2} \left( {\rm d}\theta^2 
        + \sin^2\theta\,{\rm d}\phi^2 \right) \Big] 
   + \frac{4rm^2}{r+2m}
 \Big( 
{\rm d}\psi + \cos\theta\,{\rm d}\phi\Big)^2
\end{equation}
The anticommutativity of the Dirac operators $D$ and $D_f$ holds when
the dual manifold has antisymmetric torsion given by
\begin{equation}
T=-\frac{4r^2 m^2 \sin \theta}{(r+m)^2}\, {\rm d}\theta\wedge {\rm
d}\phi\wedge {\rm d}\psi\, .
\end{equation}
In fact the results for the three contributions~(\ref{bulk}), 
(\ref{boundary}) and~(\ref{chern}) 
to the index of the
dual Dirac operator $D_f$ are given by each of the following
three lines respectively~\cite{kas_index}
\begin{eqnarray}
{\rm index}D_f&=&  
\frac{r_b^8+12r_b^7m+86r_b^6m^2+340r_b^5m^3+753r_b^4m^4+872r_b^3m^5+408r_b^2m^6}
{12(r_b+2m)^4(r_b+m)^4}\nonumber\\
&&\!\!\!\!-\,
\frac{r_b^8+12r_b^7m+86r_b^6m^2\!+308r_b^5m^3\!+569r_b^4m^4\!+552r_b^3m^5\!+264r_b^2m^6\!+48r_bm^7}
{12(r_b+2m)^4(r_b+m)^4}\nonumber\\
&& \!\!\!\!\!\!\hphantom{-\,r_b^8+12r_b^7m+86r_b^6m^2}
-\,\frac{32r_b^5m^3+184r_b^4m^4+320r_b^3m^5+144r_b^2m^6-48r_bm^7}
{12(r_b+2m)^4(r_b+m)^4}
\nonumber\\
&=&\,0 \, .
\end{eqnarray}
Note that the vanishing result holds for any positive radius to the boundary
$r_b$ and this is a very strong test of the formalism presented.
Let us conclude by remarking 
that in theories with local supersymmetry, torsion is
inevitably present. Furthermore boundary physics in these theories has
become increasingly important~\cite{mald2}, a fact which highlights the
interest of our work.

\vs{2} 

\nit
{\bf Acknowledgement} 
For two of us (J.W.v.H.\ and A.W.) this work is part of the 
research programme of the Foundation for Fundamental Research of
Matter (FOM).  

\begin{small}
\bibliographystyle{utphys}
\bibliography{letterversion}
\end{small}

\end{document}